\documentclass[12pt, amsmath,amssymb]{revtex4}
\usepackage{graphicx}
\usepackage{dcolumn}
\usepackage{bm}
\usepackage{wrapfig}
\usepackage{amsmath,amsfonts,amssymb,bm,ulem}
\usepackage[usenames]{color}
\begin{document}

\title{Hall detection of a ppm spin polarization}

\author{Dazhi Hou$^{1,2,\star}$, Z. Qiu$^{1}$, R. Iguchi$^{3}$,  K. Sato$^{1}$, K. Uchida$^{2,3,4}$, G. E. W. Bauer$^{1,3,5}$, E. Saitoh$^{1,2,3,6}$}
\affiliation{
$^1$ WPI Advanced Institute for Materials Research, Tohoku University, Sendai 980-8577, Japan\\
$^2$ ERATO, Japan Science and Technology Agency, Sanbancho, Tokyo 102-0076, Japan \\
$^3$ Institute for Materials Research, Tohoku University, Sendai 980-8577, Japan \\
$^4$ PRESTO, Japan Science and Technology Agency, Saitama 332-0012, Japan \\
$^5$ Kavli Institute of NanoScience, Delft University of Technology, 2628 CJ Delft, The Netherlands \\
$^6$ Advanced Science Research Center, Japan Atomic Energy Agency, Tokai 319-1195, Japan \\
$^{\star}$Electronic address: dazhi.hou@imr.tohoku.ac.jp}

\begin{abstract}

\end{abstract}


\maketitle

\textbf{Hall effects have been employed as sensitive detectors of magnetic fields and magnetizations\cite{Hall1st,Nagaosa_AHEreview}. In spintronics, exotic phenomena often emerge from a non-equilibrium spin polarization or magnetization\cite{FertGMR,HirschSHE,Tserkovnyak2002,SpintronicsRevModPhys}, that is very difficult to measure directly. The challenge is due to the tiny total moment, which is out of reach of superconducting quantum interference devices and vibrating sample magnetometers or spectroscopic methods such as X-ray magnetic circular dichroism. The Kerr effect is sufficiently sensitive only in direct gap semiconductors, in which the Kerr angle can be resonantly enhanced\cite{Kato2004}. Here we demonstrate that even one excess spin in a million can be detected by a Hall effect at room temperature. The novel Hall effect is not governed by the spin Hall conductivity but by its energy derivative thereby related to the spin Nernst effect\cite{spinNerstmertig}.}

In principle, a non-equilibrium magnetization ($\widetilde{\textbf{m}}$) should induce an electromotive force $\textbf{E}_{\textrm{AHE}}$ perpendicular to an applied current density $\textbf{j}_\textrm{c}$ analogous to the anomalous Hall effect (AHE)\cite{Nagaosa_AHEreview,Fe_intrinsic_cal} in ferromagnets as illustrated in Fig. 1:
\begin{equation}\label{eq0}
\textbf{E}_{\textrm{AHE}}\propto \textbf{j}_\textrm{c}\times \widetilde{\textbf{m}}.
\end{equation}
Since Hall measurements do not require specific structures or sample sizes, they may serve as a more flexible electrical probe of $ \widetilde{\textbf{m}}$ than the contact voltage to ferromagnetic electrodes\cite{Johnson1985,Valenzuela2006,Kimura2007}. However, such a non-equilibrium anomalous Hall effect has only been observed in a few semiconductors in which the optically excited spin polarization $P$, i.e. the ratio of the spin polarized to the total density, is very high\cite{Miah2007AHEinGaAs,SIHE,Okamoto2014} ($P>$ 10 \%). In conventional metals the spin polarization due to the non-equilibrium magnetization is far below 0.1\%\cite{SpinCurrent}, which appears to have discouraged researchers to attempt detection by a Hall voltage.

In this paper we show that a non-equilibrium magnetization causes a detectable Hall voltage even in a metal and at room temperature owing to a novel mechanism formulated below. Figure 2a is a schematic illustration of our measurement setup. The sample is a bilayer of nonmagnetic iridium-doped copper ($\textrm{Cu}_{95}\textrm{Ir}_{5}$) and a ferrimagnetic insulator yttrium iron garnet (YIG). A non-equilibrium magnetization $\widetilde{\textbf{m}}$ is injected into the $\textrm{Cu}_{95}\textrm{Ir}_{5}$ layer via the spin pumping from the YIG layer\cite{Tserkovnyak2002,Kajiwara2010}, the orientation of which is antiparallel to the YIG magnetization. We apply a magnetic field normal to the sample surface so that the YIG magnetization is aligned perpendicular to the interface. In this out-of-plane magnetization configuration, there is no dc inverse spin Hall effect (ISHE) by symmetry\cite{Ando2008} (see the Supplementary Information). We then apply an in-plane electric current, $\textbf{j}_\textrm{c}$, and measure the Hall voltage, $V_\textrm{H}$, in the direction perpendicular to the current using a five-probe method. All the measurements are carried out at room temperature.

Figure 2b shows the Hall voltage in a $\textrm{Cu}_{95}\textrm{Ir}_{5}$(24 nm)/YIG sample without microwave excitation. The observed Hall voltage is proportional to the external magnetic field, as expected for the ordinary Hall effect. The inset shows a Hall resistance that linearly scales with magnetic fields up to 10000 Oe, which is well above the magnetization saturation field of the YIG layer ($<$ 2000 Oe), clearly excluding an anomalous-type Hall effect generated, e.g., in the $\textrm{Cu}_{95}\textrm{Ir}_{5}$ layer next to YIG if magnetized by an equilibrium proximity effect\cite{Huang2012}.

Figure 2c is the absorption spectrum of microwaves at 50 mW power; the data shows the ferromagnetic resonance (FMR) of the YIG layer around 4920 Oe. In Fig. 2d, we show the Hall voltage measured when applying the same microwave powers and different currents. For vanishing currents (the black curve at the bottom in Fig. 2d), the Hall voltage is almost zero even at the FMR, which confirms that the ISHE is forbidden in the out-of-plane magnetization configuration\cite{Ando2008}.

Surprisingly, when the microwave and electric currents are simultaneously applied, a conspicuous Hall voltage emerges at the ferromagnetic resonance field $H_{\textrm{FMR}}$, as shown in Fig. 2d, which scales with the applied current. The peak voltage dependence on the out-of-plane magnetic field angle $\theta_{\rm{H}}$ (see Fig. 2e) is shown in Fig. 2f;  its sign changes by reversing the field direction and it vanishes for in-plane magnetic fields ($\theta_{\rm{H}}= 0$ and $180^{\circ}$). The odd magnetic-field symmetry manifests time-reversal symmetry breaking in $\textrm{Cu}_{95}\textrm{Ir}_{5}$, which proves that a spin polarization is the origin of the observed features.

The Hall voltage measured under microwave irradiation and for currents with opposite directions is displayed in Fig. 3a. Both the ordinary Hall effect (background slope) and the peak at $H_{\textrm{FMR}}$ change their signs with the current reversal, therefore the ISHEs or magnetoelectric rectification \cite{Saitoh2006,GuiPhysRevLett.98.107602,PhysRevLett.111.217602} are clearly ruled out. The magnitude of the peak signal is proportional to the applied current, as shown in Fig. 3b, consistent with Eq. (1) and, hence, is attributed to the anomalous Hall effect, which disappears when microwave is off.

Figures 3c and 3d show that the measured anomalous Hall voltage $V_{\textrm{AHE}}$ scales proportionally with the microwave power. In sharp contrast, the normal Hall contribution does not depend on excitation power. This distinctive difference in the microwave response demonstrates that the peak signal is caused by spin pumping. We confirmed that the out-of-plane magnetization angle ($\theta_\textrm{M}$) dependence of the normalized AHE signal (open circles in Fig. 3e) is well reproduced by the perpendicular component of the non-equilibrium magnetization in the metal layer as predicted in Eq. (1) (see the Supplementary Information)\cite{Ando2008}.

Figure 3f shows the Hall voltage measured in a $\textrm{Cu}_{95}\textrm{Ir}_{5}$(12 nm)/YIG sample and in a  $\textrm{Cu}_{95}\textrm{Ir}_{5}$(12 nm)/$\textrm{SiO}_2$(10 nm)/YIG sample in which a $\textrm{SiO}_2$ layer is inserted between the $\textrm{Cu}_{95}\textrm{Ir}_{5}$ and the YIG layers. At zero microwave excitation, the both samples show similar ordinary Hall effects. On the other hand, at 50 mW microwave excitation, the AHE peak signal is absent in the presence of the $\textrm{SiO}_2$ layer because the spin pumping process requires direct exchange interaction between the magnet and the metal layer\cite{Mosendz2010MgOinsert,Du2013barrierinsert}. The observed suppression of the AHE signal by the insulating layer again supports our scenario of a non-equilibrium AHE.

We phenomenologically describe the observations using a diffusion theory of the spin Hall effect (SHE). The non-equilibrium magnetization in $\textrm{Cu}_{95}\textrm{Ir}_{5}$ can be considered due to a spin accumulation as sketched in Fig. 1, of which the magnitude is defined as the difference between chemical potentials of spin-up and spin-down electrons: $\mu_s=\mu^\uparrow-\mu^\downarrow$. Here we define the up-spin direction is antiparallel to $\widetilde{\textbf{m}}$. When a longitudinal electric field ($\textbf{E}$) is applied, Hall currents are induced via the spin-orbit interaction\cite{HirschSHE,NiimiCuIr}: $\textbf{j}_\textrm{H}^{\uparrow(\downarrow)}=\sigma_{\textrm{SHE}}^{\uparrow(\downarrow)}\boldsymbol{\sigma}\times\textbf{E},
$
where $\sigma_{\textrm{SHE}}^{\uparrow(\downarrow)}$ is the spin Hall conductivity for the up(down) spin electrons and $\boldsymbol{\sigma}$ the electron spin polarization vector. For small $\mu_s$, $\sigma_{\textrm{SHE}}^{\uparrow(\downarrow)}$ can be expanded as
$\sigma_{\textrm{SHE}}^{\uparrow(\downarrow)}=1/2[\sigma_{\textrm{SHE}}\pm(\partial\sigma_{\textrm{SHE}}/\partial\varepsilon)\mu_s/2]$, in which $\sigma_{\textrm{SHE}}$ is the spin Hall conductivity in the equilibrium state and the derivative is taken at the Fermi energy. The total SHE-induced Hall currents reads:
\begin{equation}\label{eq0}
\textbf{j}_{\textrm{AHE}}=\textbf{j}_\textrm{H}^{\uparrow}+\textbf{j}_\textrm{H}^{\downarrow}=\frac{\partial\sigma_{\textrm{SHE}}}{\partial\varepsilon}\frac{\mu_s}{2}\frac{-\widetilde{\textbf{m}}}{|\widetilde{\textbf{m}}|}\times\textbf{E},
\end{equation}
where $-\widetilde{\textbf{m}}/|\widetilde{\textbf{m}}|$ denotes the spin polarization vector of the more populated sub-band. According to Eq. (2), the non-equilibrium AHE in the linear response in $\mu_s$ is proportional to the energy derivative of the spin Hall conductivity ($\partial\sigma_{\textrm{SHE}}/\partial\varepsilon$), while the conventional spin Hall effect (anomalous Hall effect) scales with $\sigma_{\textrm{SHE}}$ ($\sigma_{\textrm{AHE}}$). Figure 4 shows the $\textrm{Cu}_{95}\textrm{Ir}_{5}$-thickness dependence of the non-equilibrium AHE resistance measured at 50 mW microwave excitation; the decay trend is well explained by Equation (2) (see Equation (S7) in the Supplementary Information). The fit in Fig. 4 yields the spin diffusion length in $\textrm{Cu}_{95}\textrm{Ir}_{5}$ $\lambda_s=24\ \textrm{nm}$, the spin mixing conductance of the $\textrm{Cu}_{95}\textrm{Ir}_{5}$/YIG interface $g_{r}^{\uparrow \downarrow}=8.4\times10^{18}/\textrm{m}^2$ and the energy derivative of the spin Hall conductivity $\partial\sigma_{\textrm{SHE}}/\partial\varepsilon= -9620\ \Omega^{-1}\textrm{m}^{-1}/\textrm{meV}$. $\partial\sigma_{\textrm{SHE}}/\partial\varepsilon$ has the  same sign as $\sigma_{\rm{SHE}}$ in $\textrm{Cu}_{95}\textrm{Ir}_{5}$\cite{NiimiCuIr} which suggests an enhanced spin Hall efficiency with increasing Fermi energy. We thus can calculate the thickness dependence of the averaged spin accumulation $\widetilde{\mu_s}$ and spin polarization $\widetilde{P}$ as shown in the inset of Fig. 4 (see the Supplementary Information for its definition)\cite{Mazinpolarization}, in which $\widetilde{P}$ is as small as 1 ppm under our experimental conditions. $\partial\sigma_{\textrm{SHE}}/\partial\varepsilon$ is closely related to the spin Nernst coefficient as calculated from first principles for various impurities in copper, i.e. $\partial\sigma_{\textrm{SHE}}/\partial\varepsilon= 2000\ \Omega^{-1}\textrm{m}^{-1}/\textrm{meV}$ for gold-doped copper and $\partial\sigma_{\textrm{SHE}}/\partial\varepsilon= 230\ \Omega^{-1}\textrm{m}^{-1}/\textrm{meV}$ for bismuth-doped copper\cite{spinNerstmertig}.

$\partial\sigma_{\textrm{SHE}}/\partial\varepsilon$ can be decomposed into two terms:
\begin{equation}\label{eq0}
\frac{\partial\sigma_{\textrm{SHE}}}{\partial\varepsilon}=\frac{\partial\sigma}{\partial\varepsilon}\theta_{\textrm{SHE}}+\frac{\partial\theta_{\textrm{SHE}}}{\partial\varepsilon}\sigma
\end{equation}
where we define the spin Hall angle as $\theta_{\textrm{SHE}}=\sigma_{\textrm{SHE}}/\sigma$ and $\sigma$ is the conductivity of the $\textrm{Cu}_{95}\textrm{Ir}_{5}$ film. Since $\partial\sigma/\partial\varepsilon=S/eL_0T$ ($S$ the Seebeck coefficient, $L_0$ the Lorenz number)\cite{ashcroft2011solid}, we may estimate $\theta_{\textrm{\textrm{SHE}}}\partial\sigma/\partial\varepsilon= - 35\ \Omega^{-1}\textrm{m}^{-1}/\textrm{meV}$ with $S= 3.5$ $\mu$V/K as measured in a $\textrm{Cu}_{95}\textrm{Ir}_{5}$ film and $\theta_{\textrm{SHE}}= -0.02$ from Ref.\cite{NiimiCuIr}. This estimate implies that the first term in Eq. (3) is negligibly small in the present system, while the second term dominates with $\partial\theta_{\textrm{SHE}}/\partial\varepsilon= -2.6$/eV. Such a strongly energy-dependent spin Hall angle, which can be semi-qualitatively understood with an independent theoretical study on resonant skew scattering of iridium impurities in copper\cite{Xu2014} that suggests similar behavior in other systems with resonant skew scattering\cite{spinNerstmertig}. While the spectral properties of the spin Hall effect can be investigated by finite bias voltages\cite{Liu2014}, $\partial\theta_{\textrm{SHE}}/\partial\varepsilon$ has not been accessed yet. The present method thus provides information relevant for designing and testing spin Hall materials\cite{MertigISHEcal}.

In conclusion, the Hall detection of ppm-order spin polarization offers a straightforward approach to detect tiny non-equilibrium magnetizations electrically and non-invasively using simple sample structures without ferromagnetic electrodes. This measurement also manages to determine the energy dependence of the spin Hall conductivity around the Fermi energy without carrier doping or thermal gradients and is applicable to a wide range of materials. Our method should help setting up an extensive database of spin Hall and spin Nernst coefficients for conductors that are potentially relevant for spin Hall devices.

\textbf{Method}

The single-crystal yttrium iron garnet (YIG) films used in the present work were prepared by liquid phase epitaxy (LPE) on a gallium gadolinium garnet substrate. All samples were cut from a single 4.5 $\mu$m LPE YIG film. To achieve good-YIG/$\textrm{Cu}_{95}\textrm{Ir}_{5}$ interfaces, the YIG films underwent an acid pickling process before being transferred into high vacuum under which they were annealed $in$-$situ$ for three hours at 500$^\circ$ C before the $\textrm{Cu}_{95}\textrm{Ir}_{5}$ layer was sputtered at room temperature. The $\textrm{Cu}_{95}\textrm{Ir}_{5}$ thickness was calibrated by X-ray reflectometry using control samples. The crystallographic properties of the YIG/$\textrm{Cu}_{95}\textrm{Ir}_{5}$ device were characterized by transmission electron microscopy (TEM) and x-ray diffractometry, to confirm that the YIG films were high quality single crystals with a lattice constant of 12.376 {\AA}, while the $\textrm{Cu}_{95}\textrm{Ir}_{5}$ layers showed multi-crystalline structures without a preferred orientation. During electric measurement the sample is placed at the center of a $\textrm{TE}_{\textrm{011}}$ microwave cavity with the resonance frequency of 9.45 GHz and the microwave field is along the x axis as defined in Fig. 2a.

\newpage

\bibliographystyle{naturemag}


\newpage
\textbf{Acknowledgement}

The authors thank Zhuo Xu and Jana Lustikova for valuable discussions. This work was supported by ERATO ``Spin Quantum Rectification'', PRESTO ``Phase Interfaces for Highly Efficient Energy Utilization'', Strategic International Cooperative Program ASPIMATT from JST, Japan, Grant-in-Aid for Scientific Research (A) (24244051, 25247056), Grant-in-Aid for Scientific Research on Innovative Area, ``Nano Spin Conversion Science'' (26103006), Grant-in-Aid for Young Scientists (A) (25707029), Grant-in-Aid for Young Scientists (B) (26790038), Grant-in-Aid for Challenging Exploratory Research (26600067), Grant-in-Aid for Research Activity Start-up (25889003) from MEXT, Japan, NEC corporation, the Casio Science Promotion Foundation, and the Iwatani Naoji Foundation.

\textbf{Author contributions}

D.H. designed the experiment, collected all of the data and analyzed the data. E.S. supervised the study. Z.Q., D.H. fabricated the samples and performed the TEM characterization of the sample. D. H., R. I., K. S., K. U., Z. Q., G. E. W. B. and E. S. developed the explanation of the experiment. D. H., R. I., Z. Q., K. S. and E.S. wrote the manuscript. All authors discussed the results and commented on the manuscript.

\newpage

\textbf{Figure Captions:}
 \\

\textbf{Fig. 1:} \textbf{The anomalous Hall effect induced by the non-equilibrium magnetization.} In the presence of a non-equilibrium magnetization $\widetilde{\textbf{m}}$, the spin-up and spin-down electrons are unequally populated in a normal metal (NM) and have spin Hall conductivities $\sigma_{\textrm{SHE}}^{\uparrow(\downarrow)}$ that depend on the spin-dependent chemical potentials $\mu^{\uparrow(\downarrow)}$. When a charge current $\textbf{\textrm{j}}_\textrm{c}$ is applied normal to $\widetilde{\textbf{m}}$, a finite anomalous Hall-like voltage may be expected.
\newpage

\textbf{Fig. 2:} \textbf{Experimental setup and observation of the non-equilibrium anomalous Hall effect.} \textbf{a}, $\textrm{Cu}_{95}\textrm{Ir}_{5}$/$\textrm{Y}_\textrm{3}\textrm{Fe}_\textrm{5}\textrm{O}_{\textrm{12}}$ bilayer structure and Hall effect measurement setup. Charge current is applied along the x axis and the Hall voltage is measured in the y direction. The magnetic field is applied normal to the sample surface which is a 2 mm$\times$3 mm rectangle. \textbf{b}, Hall voltage measured in a $\textrm{Cu}_{95}\textrm{Ir}_{5}(24 \textrm{nm})$/YIG sample at different applied currents without microwave excitation. The dashed lines are alinear fit of the normal Hall contribution. The inset shows the Hall resistance in a larger magnetic field range. \textbf{c}, Field($H$) dependence of the FMR signal ($\textrm{d}I/\textrm{d}H$) under 50 mW microwave excitation, and $H_{\textrm{FMR}}$= 4920 Oe is the ferromagnetic resonance field.  \textbf{d}, The Hall voltage under 50 mW microwave excitation with different applied currents. \textbf{e}, Definition of the magnetic field angle $\theta_\textrm{H}$ and the magnetization angle $\theta_\textrm{M}$ when applying a magnetic field with out-of-plane component. \textbf{f}, The Hall voltage measured with in-plane ($\theta_\textrm{H}=0^{\circ}, 180^{\circ}$) and perpendicular ($\theta_\textrm{H}=90^{\circ}, -90^{\circ}$) magnetic fields at 50 mW microwave excitation.
\newpage

\textbf{Fig. 3:} \textbf{Systematic measurement of non-equilibrium anomalous Hall effects.} \textbf{a}, Field dependence of the Hall voltage measured with currents of constant magnitude but opposite polarities at 50 mW microwave excitation, $|$\textit{I}$|$= 30 mA (corresponds to a very low current density of $1.15\times10^9 \textrm{A/m}^2$). \textbf{b}, The current dependence of the $V_{\textrm{AHE}}$ signal at 50 mW microwave excitation. The solid line is a linear fit. \textbf{c}, Field dependence of the Hall voltage at different values of the microwave excitation power. The applied current is fixed at 30 mA. \textbf{d}, Microwave power $P_{\textrm{MW}}$ dependence of $V_{\textrm{AHE}}$ in Fig. 3c. The solid line is a linear fit. \textbf{e}, Magnetization angle ($\theta_\textrm{M}$) dependence of $V_{\textrm{AHE}}(\theta_\textrm{M})$/$V_{\textrm{AHE}}(\theta_\textrm{M}=90^\circ)$. The filled circles are experimental data. The solid curve represents a model calculation without fitting parameters (see Eq. (S11) in the Supplementary Information). \textbf{f}, Field dependence of the Hall voltage measured in a $\textrm{Cu}_{95}\textrm{Ir}_{5}$(12 nm)/YIG (blue curves) and a $\textrm{Cu}_{95}\textrm{Ir}_{5}$(12 nm)/$\textrm{SiO}_\textrm{2}$(10 nm)/YIG sample (black curves). The microwave excitation power is 0 and 50 mW as labeled.
\newpage

\textbf{Fig. 4:} \textbf{Thickness dependence of the non-equilibrium anomalous Hall effect.} The non-equilibrium AHE resistance in the samples with different $\textrm{Cu}_{95}\textrm{Ir}_{5}$ thicknesses at 50 mW microwave excitation. The filled circles are the experimental data, and the solid curve is a fit based on Equation (2) (see the Supplementary Information). The inset shows the thickness dependence of the averaged spin accumulation and spin polarization in $\textrm{Cu}_{95}\textrm{Ir}_{5}$ films at different microwave excitation.
\newpage

\begin{figure}
\centering
  \includegraphics[width=12cm]{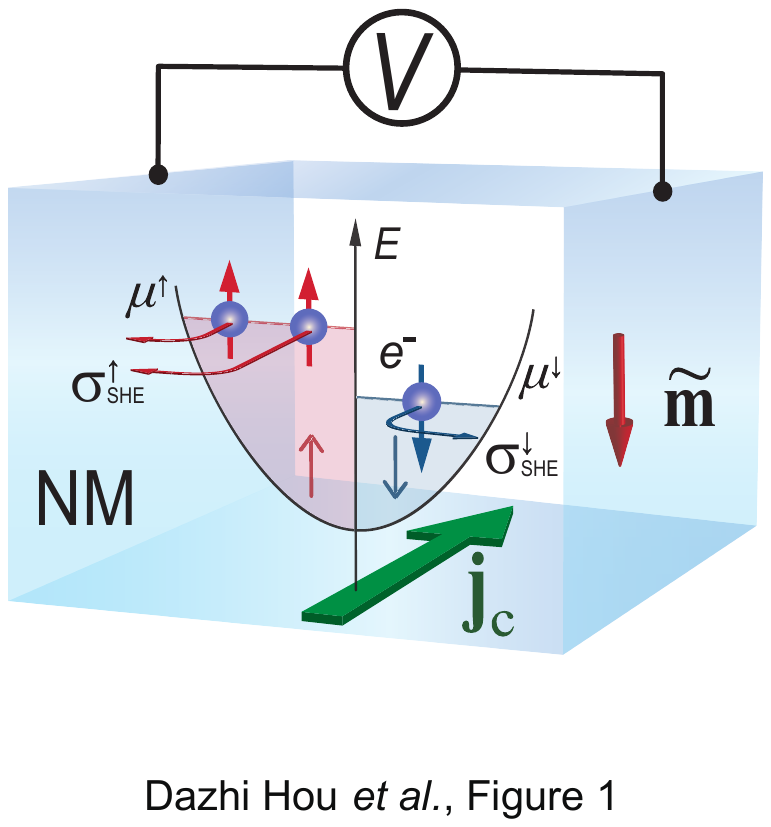}

\end{figure}

\clearpage

\begin{figure}
\centering
  \includegraphics[width=16cm]{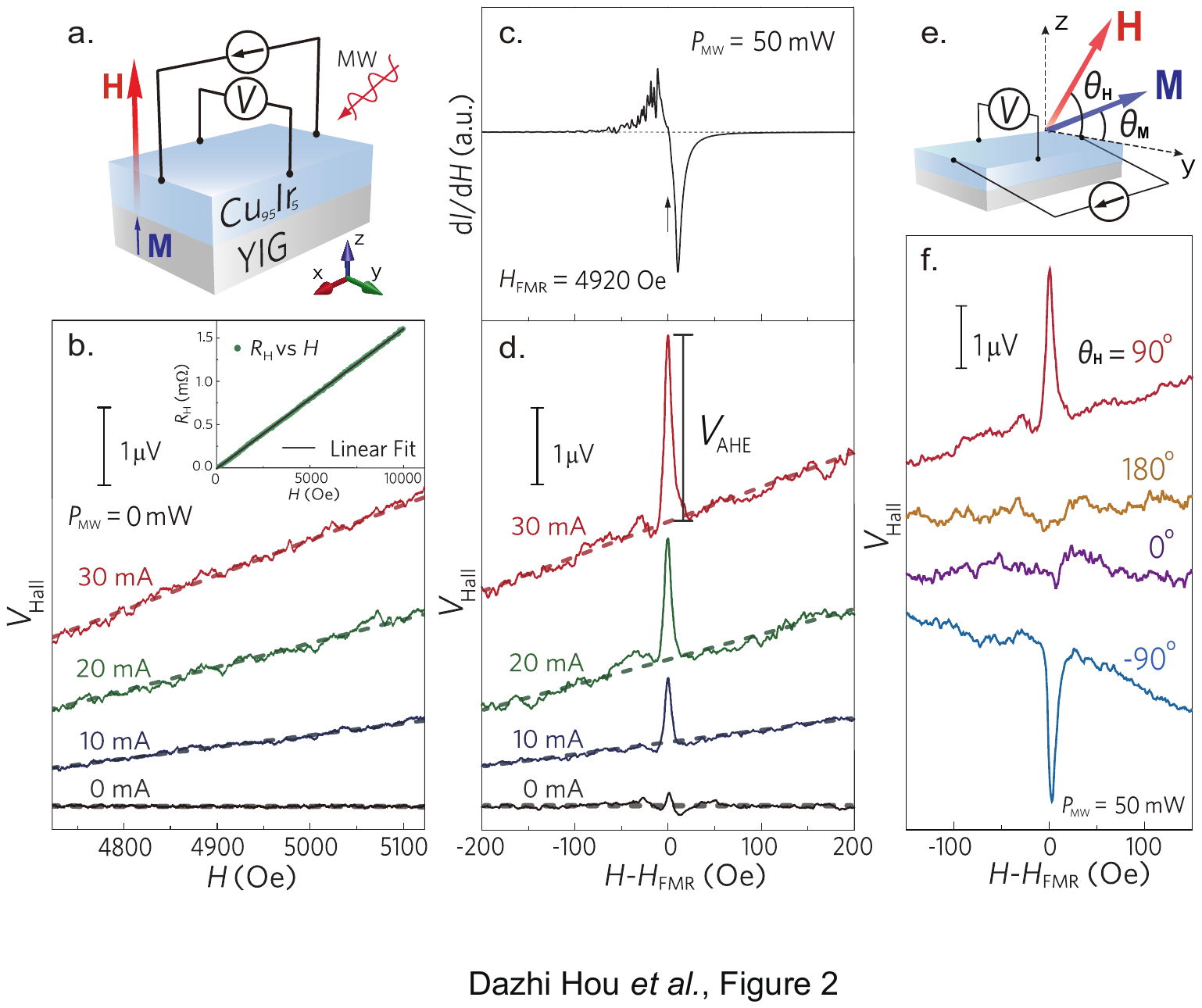}
  
\end{figure}
\clearpage

\begin{figure}
\centering
  \includegraphics[width=16cm]{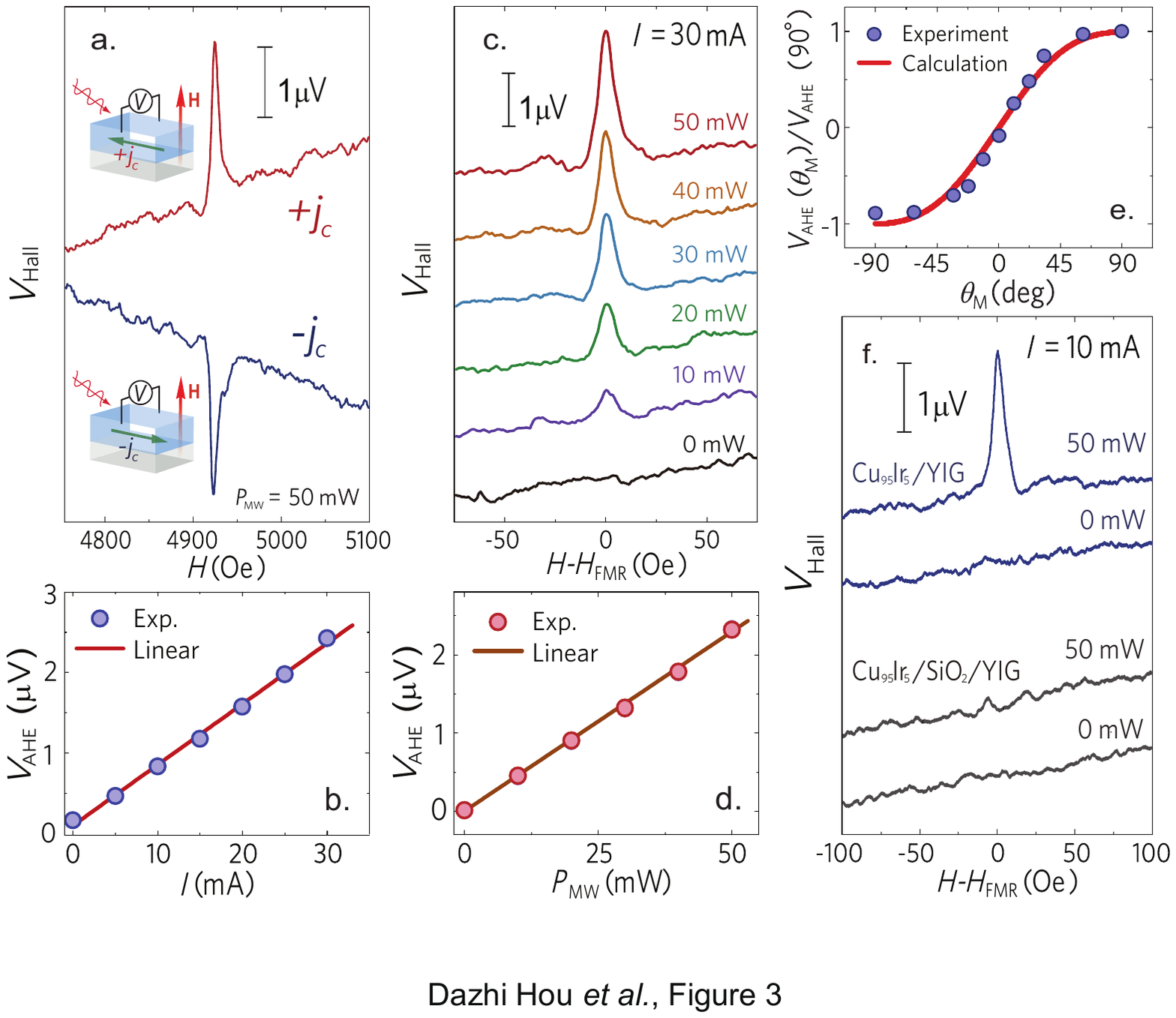}
  
\end{figure}
\clearpage

\begin{figure}
\centering
  \includegraphics[width=12cm]{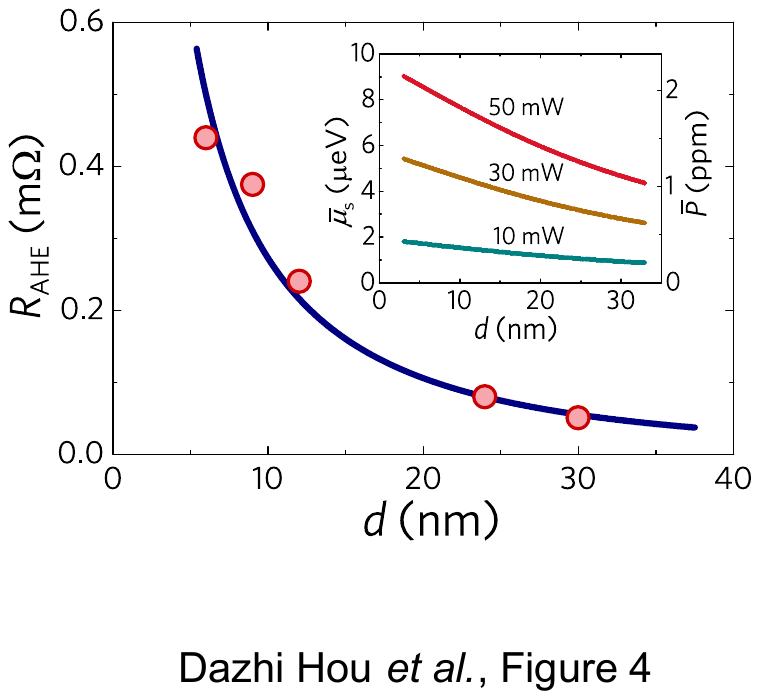}

\end{figure}

\end{document}